# Nanophotonic inspection of deep-subwavelength integrated optoelectronic chips

Ying Che[1,2,#], Tianyue Zhang [1,#,*], Xiaowei Liu[3], Dejiao Hu[4], Shichao Song[2], Yan Cai[5], Yaoyu Cao[2], Jie Zhang[1], Shi-Wei Chu[6,7,8] & Xiangping Li [2*]

[1] State Key Laboratory of Information Photonics and Optical Communications & School of Integrated Circuits, Beijing University of Posts and Telecommunications, Beijing 100876, China

[2] Guangdong Provincial Key Laboratory of Optical Fiber Sensing and Communications, Institute of Photonics Technology, Jinan University, Guangzhou 510632, China

[3] Research Center for Humanoid Sensing, Zhejiang Lab, Hangzhou 311121, China

[4] Amethystum Storage Technology Co.,Ltd, Guangzhou, China, 511493

[5]Shanghai Institute of Microsystem and Information Technology, Shanghai, 200050, China

[6] Department of Physics, National Taiwan University, Taipei 10617, Taiwan

[7] Molecular Imaging Center, National Taiwan University, Taipei 10617, Taiwan

[8] Brain Research Center, National Tsing Hua University, Hsinchu 30013, Taiwan

[#] These authors contributed equally: Ying Che, Tianyue Zhang

*Corresponding author: tianyue_zhang@bupt.edu.cn, xiangpingli@jnu.edu.cn

**Artificial nanostructures with ultrafine and deep-subwavelength feature sizes have emerged as a paradigm-shifting platform to advanced light field management, becoming a key building block for high-performance integrated optoelectronics and flat optics. However, direct optical inspection of such integrated chips with densely packed complex and small features remains a missing metrology gap that hinders quick feedback between design and fabrications. Here, we demonstrate that photothermal nonlinear scattering microscopy can be utilized for direct imaging and resolving of integrated optoelectronic chips beyond the diffraction limit. We reveal that the inherent coupling among deep-subwavelength nanostructures supporting leaky resonances allows for the pronounced heating effect to access reversible nonlinear modulations of the confocal reflection**

**intensity, leading to optical resolving power down to 80 nm (∼ $\lambda/7$). The versatility of this approach has been exemplified by direct imaging of silicon grating couplers and metalens with a minimum critical dimension of 100 nm, as well as central processing unit (CPU) chip with 45 nm technology, unfolding the long-sought possibility of in-situ, non-destructive, high-throughput optical inspection of integrated optoelectronic chips and nanophotonic chips.**

**Introduction**

Subwavelength artificial nanostructures featured with exotic optical and electronic properties, enable multi-dimensional manipulations of light field in amplitude, phase, polarization, and angular momentum[1-4], making them the vital components for implementing the next generation of integrated optoelectronic chips and flat optics. In this context, the inspection and metrology of ultrafine nanostructures densely arranged in such integrated chips completes the closed loop and provides the direct feedback mechanism to control the fabrication quality of sophisticated functional devices, which is of critical importance. Nevertheless, accessing ultrafine features at large scales has remained a daunting challenge owing to low-throughput and destructive characterizations of the existing prevailing methods like electron microscope and atomic force microscope[5]. Although incredible success achieved by ptychographic X-ray tomography, the requirement of synchrotron radiation sources is not easily accessible[6]. Therefore, the community anxiously awaits good solutions that can image large areas quickly, nondestructively and yet provide node-level high-resolution images[7-11].

Optical far-field microscopy therefore has increasingly gained significant attentions in the nanostructure characterization in materials science[12]. Innovative methods have been witnessed including taking advantage of nonlinear light-matter interactions to surpass the diffraction limit of optical imaging. Various optical nonlinearity mechanisms have already been explored and applied successfully to super-resolution imaging of graphene[13-15], nanodiamonds[16], metallic nanostructures[17-20], and semiconductive silicon[21-25] Among them, photothermal effect has been known as a ubiquitous property widely existing even in materials with a size down to a few nanometers[26]. Once subwavelength nanostructures support geometry-dependent resonance modes that can be photothermally accessed to introduce diverse nonlinear modulation of their scattering intensities, shrinking of the lateral feature sizes of the imaging of dispersed nanostructures for high-contrast and high-resolution imaging become possible[19, 20, 23, 24]. However, photothermal nonlinear scattering has not been applied to integrated optoelectronic chips with densely spaced deep-subwavelength nanostructures. Notably, photonics devices and chips leverage diverse optical resonances of the nanostructures along with the precisely patterning subwavelength structures into arrays to realize the light manipulation functionalities and improve the device efficiency[3]. The closely spaced semiconductive or metallic nanostructures inherently involve the resonant modes coupling which exhibit strong impact on their overall optical responses. The intrinsic coupled resonant modes, resonant field enhancement and nonlinear scattering properties associated with ensembles of nanostructures would be valuable to be considered in the high-resolution optical imaging process, which have rarely been considered in previous studies.

Leveraging the inherent coupling among densely patterned deep-subwavelength nanostructures supporting leaky resonances, in this study, we demonstrate that photothermal nonlinear scattering microscopy is applicable to optical resolving integrated chips well below the diffraction limit. The coupled leaky resonances lead to significantly enhanced photothermal heating effects that allow to access thermo-optic spectral shift for diverse optical nonlinear scattering, including sub-linear and super-linear intensity-dependent reflection. Therefore, it produces distinctive nonlinear point spread functions (PSFs) in the confocal laser scanning imaging. By exploiting the unique PSFs, we demonstrate super-resolved optical imaging of various integrated optoelectronic architectures, including Si grating couplers with a minimum critical dimension of 80 nm and finest periodic separation of 100 nm, as well as 45 nm node integrated chips, shedding new light on the contactless and label-free sub-diffraction optical inspection of nano circuits and integrated nanophotonic chips.

**Results**

The principle of photothermal nonlinear scattering imaging is schematically illustrated in Figure 1. Without loss of generality, subwavelength-sized rectangular Si nanostructures are considered, which are the fundamental elements used in optical antennas, metamaterials and optoelectronic devices. The optical resonances, termed Mie or leaky-mode resonances[27] excited inside the Si nanowires (Si-NWs), including both electric and magnetic resonant modes, are reported to play dominant roles in governing the optical properties such as absorption and scattering. For most practical applications, the NWs are arranged in structural arrays, as presented in Fig. 1b, pushing the studies from the single-particle level to the microscopically engineered, upscaled devices. High filling fractions and large active surface demand close spacing of nanostructures to fully develop the performance and the efficiency[28]. The general existence of leaky-mode mechanisms in the nanostructure arrays yields various appealing applications in the perfect light absorber[29], solar cells and photodetectors [30], optical modulators[31, 32], enhanced light emission and lasing[33], enhanced nonlinear harmonic generation[34] and integrated photonics[35]. In the current study, the potentials of leaky modes coupling are further unveiled to benefit the optical microscopy enabling the enhanced imaging resolution.

Silicon behaves as an absorptive system upon the irradiance by the continuous-wave (CW) visible laser light. Optical coupling may provide a promising means to enhance the light absorption in deep-subwavelength NW arrays although the individual Si-NW may produce weak optical heating. Once the NWs are heated by the excitation laser beam, the leaky modes red-shift due to the thermo-optical effect[36-40] in silicon, thereby modifying the reflectance of the arrays. As shown in Figure 1c, for a given laser wavelength of interest, the Si-NW array with different pitch sizes and the wire widths might support the leaky modes with certain detuning from the laser wavelength. Therefore, the power-dependent reflection relies on the position of the resonance relative to the laser wavelength, inducing the reflectance increasing or decreasing with the growth of the laser power, yielding the super-linear or sub-linear responses, respectively. Meanwhile, such photothermal nonlinearities are directly linked with the effective PSFs in the confocal scanning imaging (Figure 1d), offering the potentials for far-field optical observations of nanostructured Si with an improved resolution, as the simulated super-localization (SL) results shown in Figure 1e. In the following, we will illustrate that the photothermal nonlinearity-based imaging modality presents a general approach that is suitable for the inspection and metrology of semiconductor and optoelectronic nanometric systems in the ambient environment, without any add-on apparatus on the commercial optical microscopes.

We consider one-dimensional (1D) rectangular Si-NWs with height of 220 nm and width ~ 100 nm, which is the most of interest building blocks used in the photonic integrated circuits with standard manufacturing processes. The absorption efficiency $Q_{abs}$ (defined as the ratio of the absorption cross-section with respect to the geometrical cross-section) of individual isolated Si-NWs with different widths are first analyzed rigorously by employing the finite-difference time-domain (FDTD) method, as shown in Figure 2a. It can be seen that even NWs with deep-subwavelength width of 50 nm still exhibit optical resonance effects. Extensive investigations have revealed that the observed absorption peaks originated from the excitation of leaky mode resonances, with the peak locations perfectly consistent with the specific leaky mode positions[41, 42]. The leaky modes could be further enhanced by constructing the isolated NW into a NW array. For quantitative comparison of the absorption enhancement, $Q_{abs}$ values of the individual Si-NW embedded in arrays were also calculated and plotted in Figure 2b, exemplifying that Si-NW array with filling fraction of 0.5 (i.e. the width of the NW equals to the separation distance) manifest the enhanced absorption especially in the lossy spectral regime of Si (see Supplementary Note 1 and 2 for simulation details). The analysis by the coupled-mode theory[28, 42] (CMT) reveals that the coupling between these closely-placed NWs balances the radiative and non-radiative losses, leading to substantially enhanced absorption (Supplementary Note 3).

In addition to the optical absorption, the leaky modes play dominant roles in governing the scattering of the single Si-NW as well as the reflection/ transmission of the NW arrays. As manifested in Figure 2c, the Si-NW array feature a plethora of resonances in reflection spectra depending both on the illumination wavelength and NW structural sizes, thus providing possibility for abundant nonlinear light reflection behaviors when the nanostructures absorb energy from the impinging light. The optical heating of the Si with increasing laser power induces the refractive index change due to temperature elevation, thus shifting the resonance to higher wavelengths. As we have described before, the nonlinear reflections are decided by the position of the initial resonance peak relative to the excitation wavelength (532 nm in the present study). In order to better predict the photothermal nonlinear responses, we took the derivative on the reflection spectra of the array structures of Figure 2c, over the wavelength at 532 nm, and plotted the curve in Figure 2d. A positive differential slope means the decrease in reflectance detected at 532 nm after the spectral redshift as temperature rises, indicating the trend of sub-linearity; a negative differential slope means that the reflectivity increases, representing the trend of super-linearity. It is important to note that, for an individual isolated nanostructure, the corresponding nonlinear behavior is strongly determined by its photothermally tunable scattering spectra[24, 43]. When it is structured into a periodic array, the reflection (or transmission) spectrum of the array may change greatly, indicating significantly different nonlinearity from that of a single one (Figure 2e and f). Therefore, we cannot directly relate the nonlinear photothermal behavior of a single isolated structure to that of an array structure.

To validate our analysis, we fabricated Si-NWs based on the standard silicon-on-insulator (SOI) platform and performed the photothermal nonlinearity measurements. A schematic of the optical configuration based on the standard confocal microscopy is shown in Figure 3a. A 532 nm CW laser was used to induce the light absorption and the subsequent photothermal nonlinearity. The incident light was focused on the sample surface by an air objective with a numerical aperture (NA) of 0.8, with a half wave plate to convert the linearly-polarized 532 nm light into either transverse-electric (TE) and transverse-magnetic (TM) polarizations. The rectangular cross section of NW is 200 nm wide by 220 nm high. The SEM image in the inset of Figure 3b shows that the fabricated periodic Si-NW arrays have a pitch size of 400 nm with filling fraction of 0.5. Measured reflection spectra of such array sample under the TM and TE illumination are plotted in Figure 3b and 3c, respectively, showing perfect agreement with the corresponding simulation results. The differential slope trends at 532 nm of the two spectra are opposite, indicating that opposite nonlinear behaviors are expected under these two polarizations with the same NW array. This is unambiguously verified by our experimentally measured nonlinear reflection intensity dependence on the excitation intensities. As demonstrated in Figure 3d and 3e, at low excitation intensities, the reflection signals in both polarization situations follow the linear trend. When the excitation reaches the certain threshold, the reflection intensities deviate from the linear growth, showcasing the appearance of the super-nonlinearity and sub-nonlinearity, respectively, exactly as we expected.

The photothermal nonlinearity provides an appealing opportunity for super-resolution imaging of the Si-NW arrays, since the Abbe resolution limit is formulated for linear optical system, and the consequence of nonlinearity is to modify imaging PSFs in laser-scanning microscopy[24, 44, 45]. Unlike previous studies on the photothermal imaging of individual structures, our present work goes beyond the characterization of the optical nonlinearities of a single nanostructure and to directly capture the photothermal nonlinearity and corresponding effective imaging PSFs in the arrayed system in which resonance interplay and coupling would completely alter the optical properties of the individual constitutive elements.

Given a certain incident intensity and Gaussian focal distribution of the illuminated beam, the effective nonlinear PSFs of the NW array in the confocal reflectance can be calculated via the full power-dependent reflection curves as nonlinear reflection responses. Based on the above measured nonlinear curves in Figure 3d and 3e, we are able to deduce the effective nonlinear PSFs of the NW array (see the Supplementary Note 4). With such super-liner and sub-linear effective PSFs in the photothermal nonlinear confocal microscopy, the scanning images correspond to the convolution of effective nonlinear PSFs with the densely-spaced NWs, if we consider the subwavelength-sized NWs are one-dimensional line sources composed of point-like scatters. The full-width-half-maximum (FWHM) of the PSF becomes narrow for the super-linear case; while for the sub-linear case, the PSF is broadened and exhibits a valley in the center. Nevertheless, in both cases, the convolution with effective nonlinear PSFs enables the resolution improvement of imaging, whilst the sub-linear reflection yields a negative contrast in the raw scanning image (Supplementary Figures S6-8).

Leveraging the PSF shaping empowered by photothermal nonlinearity, we experimentally demonstrate the potential for far-field optical localization of Si-NW arrays. The confocal laser scanning microscope we used is equipped with a 532-nm laser and a 0.8 NA objective lens, i.e. the diffraction limit is around 400 nm, hence our Si-NW array sample of 400 nm pitch size reaches the critical condition that is barely distinguished (Figure 3f). With the light intensity gradually increasing, it enters into the nonlinear imaging region, featuring higher imaging resolution. TM and TE polarized light were employed to scan and image the same area of the array sample, and the SL images in the x-y plane were reversed due to the distinctive nonlinear behaviors under these two polarizations. That is, the positions of the bright lines under TM polarization correspond to that of the dark lines in the TE image, and vice versa. More evidences of the reversed image are presented by the y-z views in Figures 3g and 3h, further clarifying the nonlinear behaviors of the array. To demonstrate the exciting potentials of SL imaging of multi-wires spaced well below the diffraction limit, we perform the optical localizing observation with Si-NW arrays having widths of 100, and 80 nm with corresponding pitch sizes of 200, and 160 nm, as shown in Figures 3i and 3j. We undoubtedly demonstrate that all the NWs can be distinctly identified (see Supplementary Note 5 for more results). The cross-sectional profile is plotted to show the FWHMs. Thus, it is evident that we have achieved an experimental resolution of 80 nm ($\lambda/7$) using a standard laser scanning microscope, relieving the complexity and stability issues associated with optical instrumentation for typical super-resolution inspection.

Our study on Si-NW arrays exemplifies that the inherent coupling effect among periodic nanostructures with deep-subwavelength features contributes substantially to the photo-induced heat generation and nonlinear reflection behaviors for super-localization. In particular the absorption happens universally in light-matter interactions, hence allowing a general mechanism for label-free detection of semiconductor photonic structures. Featuring ease of mass fabrication, rectangular NWs are predominantly used as building blocks in many integrated optoelectronic circuits and chips. For instance, grating couplers, wavelength filters and photodetectors compose of closely spaced rectangular nanostructures. To demonstrate our coupling-enhanced photothermal optical imaging is readily applied to label-free localization and inspection of silicon nanostructures, as a proof of principle, we imaged two architectures of Si grating couplers commonly used in integrated photonic chips. For the silicon-based chirped grating shown in Figure 4a, it has subwavelength fine features and nonuniform gap intervals. The confocal image at low excitation intensities (referred to as linear imaging) only resolves the structures with periods above the diffraction limit, while the nonlinear photothermal imaging enables to clearly distinguish two NWs with gap interval of 100 nm. We further show that the optical localization technique also works well for curved grating couplers in Figures 4g-n (see Supplementary video for more results and repeatability check). Comparison of SEM and optical images unambiguously correlates the nonlinear reflection images with grating couplers morphology. Noteworthily, during the whole imaging processes, the irradiance range is far from thermal deformation of Si structures and the localization imaging is highly reproducible.

We further show in Figure 5 that the SL principle provides an exciting pathway towards the metrology and inspection of integrated optoelectronic chips, which can be extended to many other systems such as metasurfaces as well as metallic nanostructures in the integrated circuits. Figures 5a-f show that for a large-area metalens composed of Si cylinders patterned in close vicinity on the SOI substrate, the conventional confocal imaging hardly discriminates fine features, whilst the photothermal nonlinear imaging clearly localizes and identifies individual Si cylinders. We also applied the SL technique to examine the dense areas of the integrated chip, an Intel Core processor manufactured with 45 nm technology. For densely arranged metal structures, in order to avoid sample damage caused by heat accumulation, we employed the oil immersion on the sample beneficial for heat dissipation[43], and performed the SL imaging with the corresponding objective lens with NA of 1.4. From the zoom-in view in the SEM (Figure 5i-j), the smallest feature sizes and intervals are around 65-75nm, which are severely blurred under conventional confocal imaging. Using the nonlinear photothermal imaging, all the closely arranged wires can be resolved, and the intensity cross-sections show the measured FWHM of the metal wires reaches below 60 nm. We want to remark here that the inherent potential of the optical nonlinearity is theoretically boundless if higher nonlinearity such as nearly vertical super-linear power dependency[46] could be accessed, and the resolution of nonlinear imaging techniques is unlimited. Yet practical considerations such as detrimental degradation of samples at high power densities must be taken into account.

**Discussion**

In summary, our investigations reveal that the inherent coupling between deep-subwavelength sized nanostructure arrays supporting leaky resonance can be used to access enhanced photothermal nonlinear scattering, and hence enable the optical inspection of integrated optoelectronic chips beyond the diffraction limit. The underlying mechanism is summarized as the following. The enhanced laser-induced heat generation empowered by mode coupling enables access to nonlinear modulations of the reflection intensity from NWs at mild incident light intensities on the order of MW/cm$^2$. In addition, the array structures possess rich spectral characteristics through coupling effects, which can alter the photothermal nonlinear scattering behavior of isolated structures. The resulting reversible modulation of reflection from the structure arrays helps to beat the diffraction limit through the modifications in the nonlinear PSFs of the scanning imaging system. Leveraging the distinctive PSFs, we demonstrate high-resolution localization of various architectures, including curved/chirped grating couplers, metalens and even CPU chips, showing the capability of high-precision structure identification. Our results may benefit the applications of label-free sub-diffraction microscopy on the inspection and metrology of integrated optoelectronic chips.

**Data and materials availability**

All data and materials are available from the authors upon reasonable request.

**Acknowledgements**

The authors thank Tianjin H-Chip Technology Group Corporation for assistance in sample fabrications. The authors acknowledge financial support from the National Key R&D Program of China (2021YFB2802003, 2022YFB3607300), the Fundamental Research Funds for the Central Universities (2023RC87), National Natural Science Foundation of China (NSFC) (62075084, 62325503, 62005104). SWC acknowledge funding from Taiwan National Science and Technology Council (NSTC112-2119-M-002-022-MBK, NSTC112-2321-B-002-025, NSTC112-2112-M-002-032-MY3) and the Featured Areas Research Center Program within the framework of the Higher Education Sprout Project funded by the Ministry of Education, Taiwan (NTU-112L8809).

**Author contributions**

X.L. and T.Z. conceived the idea and supervised the project. Y.C and T.Z. set up the optical system and conducted the experiments. X.L., Y.C and T.Z. analyzed data, prepared the figures, and wrote the manuscript, assisted by S.W.C. D.H assisted modelling and simulations of coupled-mode theory. X.L. and Y.Y.C. contributed to analysis of point spread functions of nonlinear reflection images. S.S supported the design of grating couplers. All the authors participated in the discussion and manuscript writing.

**Competing financial interests:** The authors declare no competing financial interests.

**Figure legends**

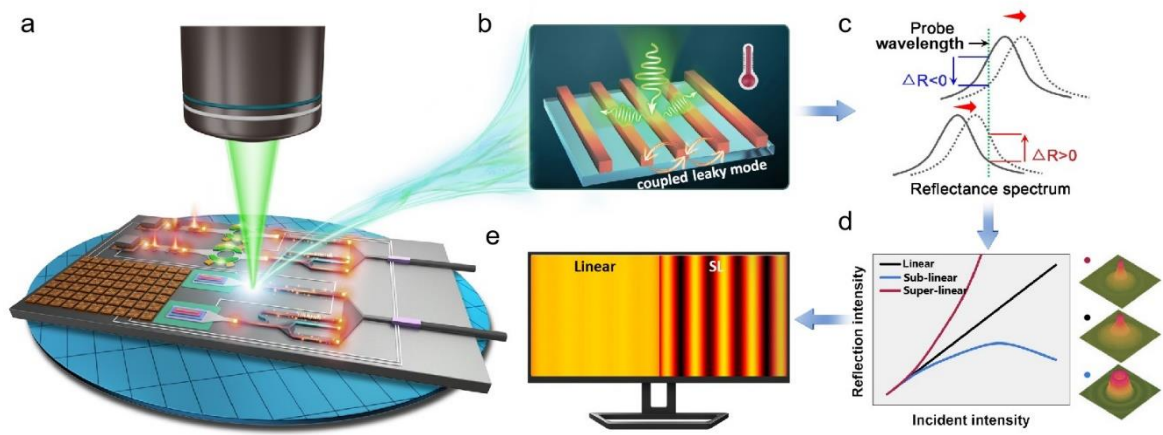

**Fig. 1. Schematic illustration of coupling-enhanced optical heating and nonlinear photothermal sub-diffraction microscopy on semiconductive nanostructures of optoelectronic integrated chip.** (a) Artistic manifestation of label-free high-resolution observation of semiconductive nanostructures of integrated photonic circuits using nonlinear photothermal imaging. (b) Illustration of coupled leaky modes enhanced optical heating of subwavelength resonant Si NW arrays. (c) When the light energy is absorbed by the sample, the refractive index changes from the thermal effect induce the shifting of the resonance frequencies. For a fixed probe wavelength, the shifts of resonances can lead to two opposite changes of reflectance ΔR, yielding the super-linearity and sub-linearity. (d) Schematics of arbitrary excitation-power-dependent reflection intensity curves for linear, super-linear and sub-linear optical responses, and the corresponding point spread functions. (e) The simulation results of imaging five NWs in close vicinity within the diffraction limit. The super-resolved localization images (denoted as SL) were attained with nonlinear reflection responses taken into account, whilst these NWs are undistinguishable in the conventional confocal imaging at the low excitation intensity (denoted as Linear).

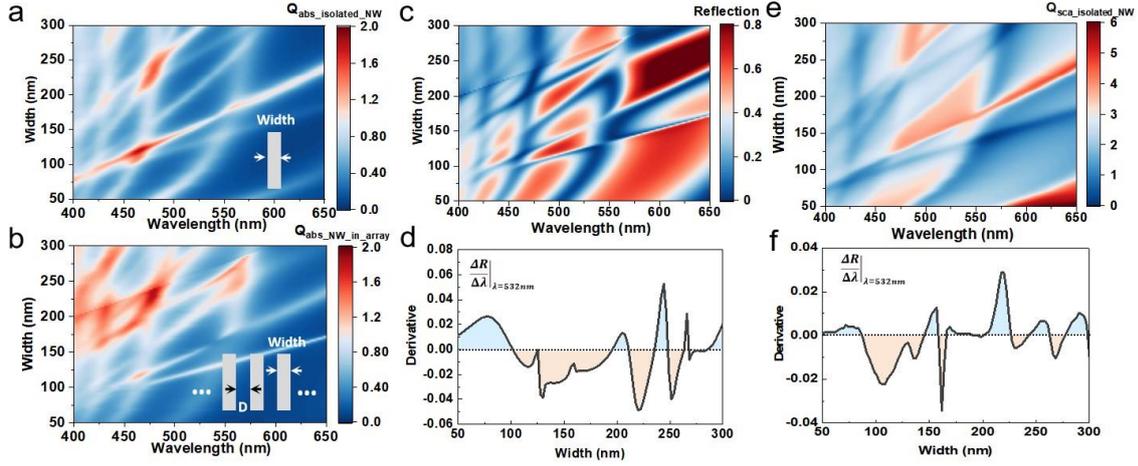

**Fig. 2. Coupled leaky modes in the Si-NW array for photothermal nonlinear reflection modulations.** (**a**) Two-dimensional color maps of absorption efficiency $Q_{abs}$ of an isolated NW as a function of the illumination wavelength and NW width under TM excitation. (**b**) The corresponding $Q_{abs}$ of the individual NWs embedded in arrays with filling factors fixed at 0.5 and increasing the NW width from 50 nm to 300 nm. (**c**) FDTD- calculated reflectance map for the Si-NW arrays (the same arrays in b) under TM excitation. (**d**) The differentiation of the reflection spectra over the wavelength at 532nm. The resulting derivatives have positive and negative values, indicating the sub-linear and super-linear behaviors both occur depending on the structural parameters. (**e**) Scattering efficiency $Q_{sca}$ of the isolated NWs with different widths. (**f**) The differentiation of the scattering spectra at 532nm. The difference between the results in figures (**d**) and (**f**) shows that the array structures and the isolated ones will exhibit different behaviors in terms of photothermal nonlinearity.

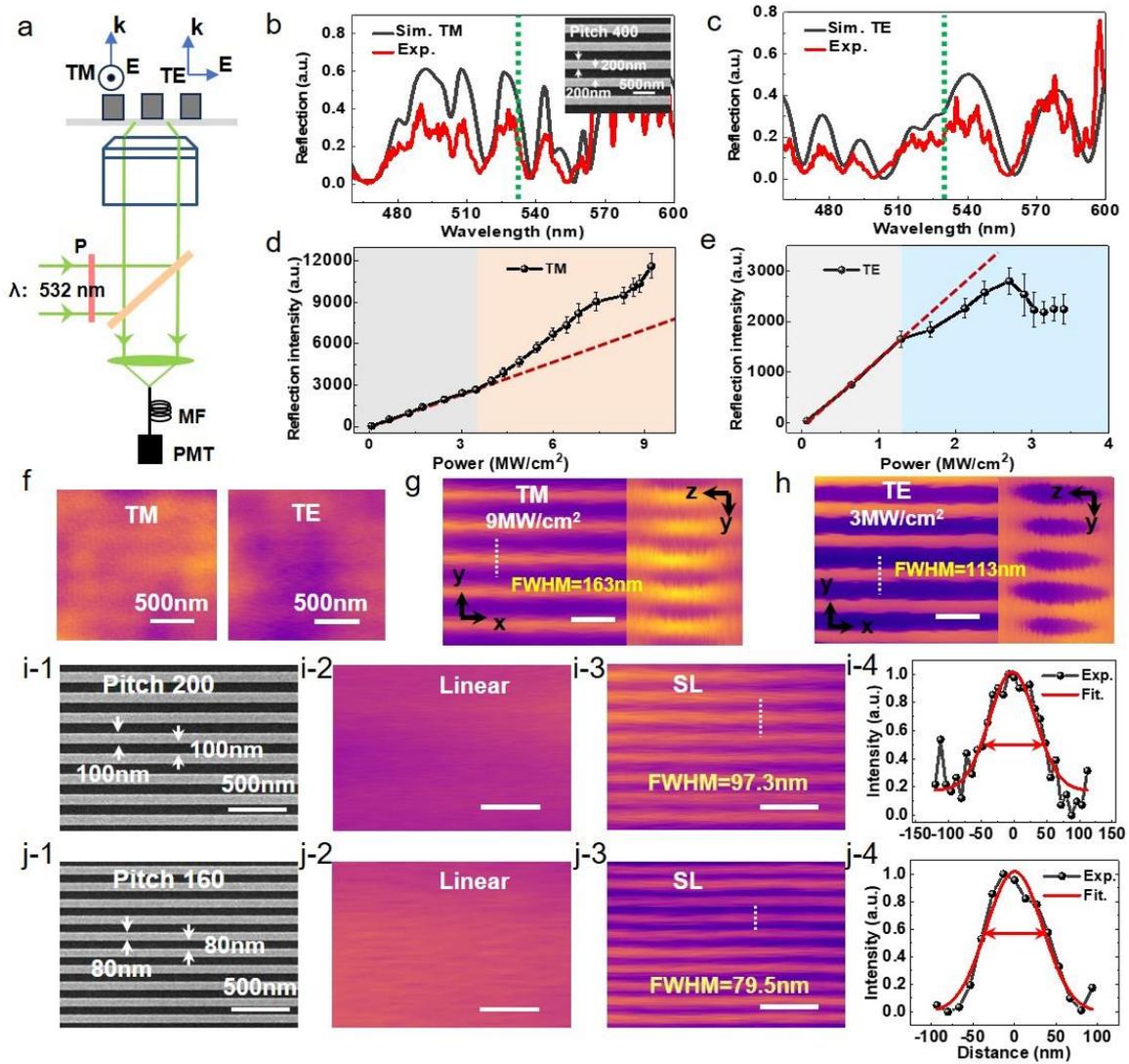

**Fig. 3. Optically localizing Si NWs arranged in dense arrays based on the photothermal nonlinearities.** **(a)** Optical setup of the reflected laser scanning confocal microscope. P: halfwave plate, BS: beam splitter, OL: objective lens, MF: multimode fiber, PMT: photomultiplier tube. Linear polarization excitations parallel (TM) or perpendicular (TE) to the NW long axis are also shown. **(b, c)** Simulated and measured reflection spectra of the array under the TM- and TE-polarized normal incidence. The inset in panel (b) is the scanning electron microscope (SEM) images of the fabricated Si-NW array with pitch size of 400 nm (width=D=200 nm). **(d, e)** The measured nonlinear dependencies of reflection intensity on irradiance intensities for the same NW array under TM-/TE-polarized excitations at the wavelength of 532 nm. The linear reflection response when the excitation intensity is low is denoted as the dash lines. For super-linearity, straightforward resolution enhancement is achieved through the PSF shrinking. Sub-linearity also leads to significant resolution enhancement providing a negative image contrast. (**f-h**) Microscopic confocal images of the 400-nm-pitch arranged Si-NW array recorded at different excitation intensities. Scale bar: 500 nm. For both polarizations at low excitation intensity, NWs are barely distinguishable. At higher intensities, photothermal nonlinearity enhanced optical imaging demonstrates resolution improvement. Images from the yz- planes in **g** and **h,** further clarifying the opposite nonlinear behaviors of the array under the two polarization excitations. More super-localization imaging results for array samples of pitch sizes of 200 nm **(i)** and 160 nm **(j)**.

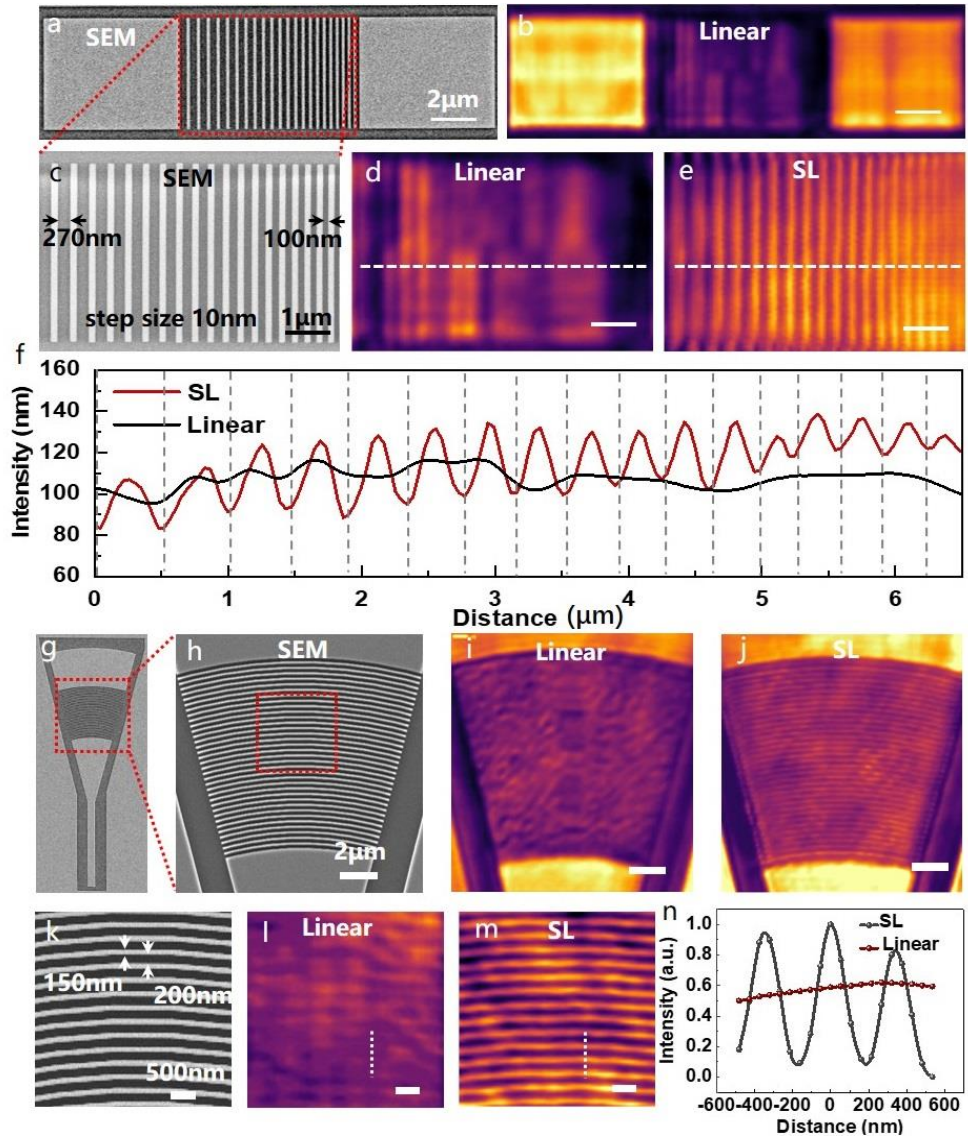

**Fig. 4. Super-localization imaging grating couplers based on the photothermal nonlinearity**. **(a)** SEM image of a nonuniform chirped grating coupler and the corresponding conventional confocal image **(b)**. **(c-e)** Zoom-in view of correlated SEM image, linear image and the photothermal nonlinear image. **(f)** The intensity cross-section profile of the selected wires (white dashed lines in **d** and **e**). **(g-n)** SEM images and the corresponding linear and nonlinear localization image of a curved grating coupler (scale bars indicated). The curved grating has duty cycle around 350 nm which is below the diffraction limit. Therefore, linear confocal imaging cannot distinguish the fine structures as shown in **(i)** and **(l)**. In contrast, far-field optical localization by means of photothermal nonlinear imaging clearly resolve these line features shown in **(j)** and **(m)**.

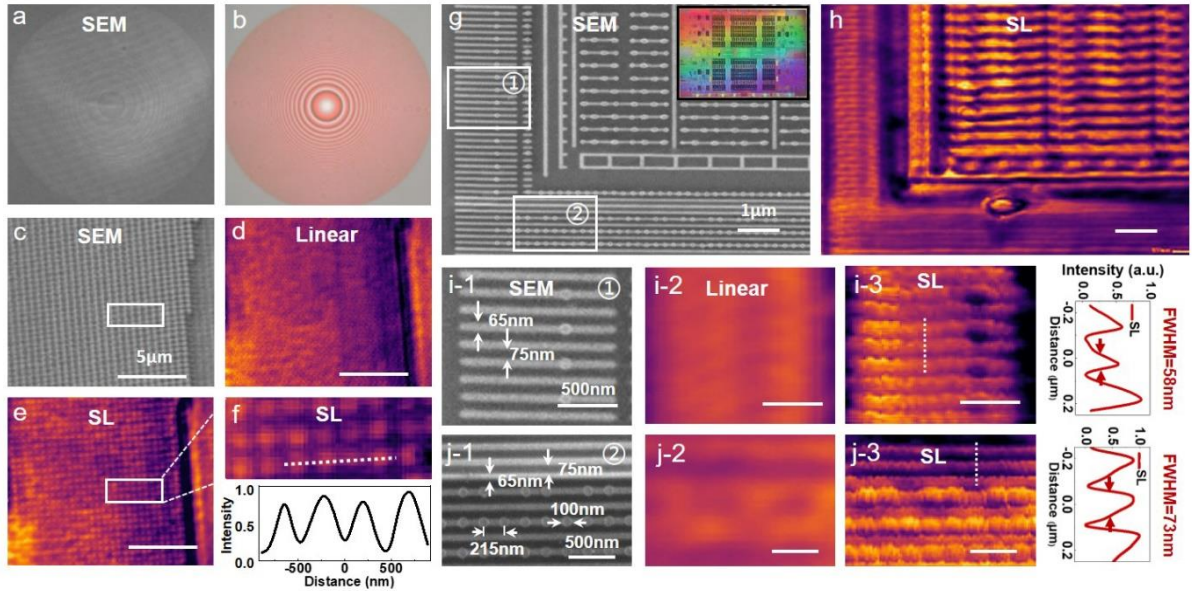

**Fig. 5. Super-localization imaging of metalens and integrated chips**. **(a)** Large-scale view of SEM image of a metalens device and the corresponding optical image **(b)**. **(c-f)** Zoom-in view of correlated SEM image, linear image and the photothermal nonlinear image. **(g-h)** Super-resolution imaging of an integrated chip. Correlated SEM image and the SL images are shown. Scale bar, 1 μm. The inset is the photograph of the chip after delayering. **(i-j)** Zoom-in images corresponding to the boxed regions in **(g)**. The measured intensity cross-section profiles of the selected wires (red dashed lines) are plotted on the right.